# CONFIDENTIAL COMPUTING IN EDGE- CLOUD HIERARCHY

Yeghisabet Alaverdyan[1,2], Suren Poghosyan[2] and Vahagn Poghosyans[2,3]

[1]EKENG CJSC, Yerevan, Armenia
yeghisabet.alaverdyan@ekeng.am

[2]Institute for Informatics and Automation Problems of NAS RA, Yerevan, Armenia
spoghosyan@iiap.sci.am

[3]Synopsys Armenia
vpoghosyan@iiap.sci.am

## ABSTRACT

*"The paper introduces confidential computing approaches focused on protecting hierarchical data within edge-cloud network. Edge-cloud network suggests splitting and sharing data between the main cloud and the range of networks near the endpoint devices. The proposed solutions allow data in this two-level hierarchy to be protected via embedding traditional encryption at rest and in transit while leaving the remaining security issues, such as sensitive data and operations in use, in the scope of trusted execution environment. Hierarchical data for each network device are linked and identified through distinct paths between edge and main cloud using individual blockchain. Methods for data and cryptographic key splitting between the edge and the main cloud are based on strong authentication techniques ensuring the shared data confidentiality, integrity and availability.*

## 1. INTRODUCTION

Edge-cloud computing architecture is still three-tier network. The very basic level in this hierarchy outlines the edge level network, where network devices and sensors reside. The second layer deploys the edge data centers which perform the operational management of individual groups of edge devices. Aimed at optimization of operational management, classification of devices prior to inclusion to some group supervised by an individual mini data center is suggested. Meanwhile, regrouping is performed according to individual groups mission and level of intelligence. At this level, data collected from sensors will be analysed and categorized before sending to the main cloud for further processing. The main cloud server is on the top of the hierarchy and still plays its own role in big data analysis, business logic maintenance and data warehousing.

The segmented architecture of the edge-cloud network enables splitting the overall workload while leaving some portion of data processing at the basic layer of the network, closer to the devices. In this sense, edge computing shouldn't be confused with autonomous computations and data processing embedded in edge devices. Figure 1 illustrates the edge computing architecture.

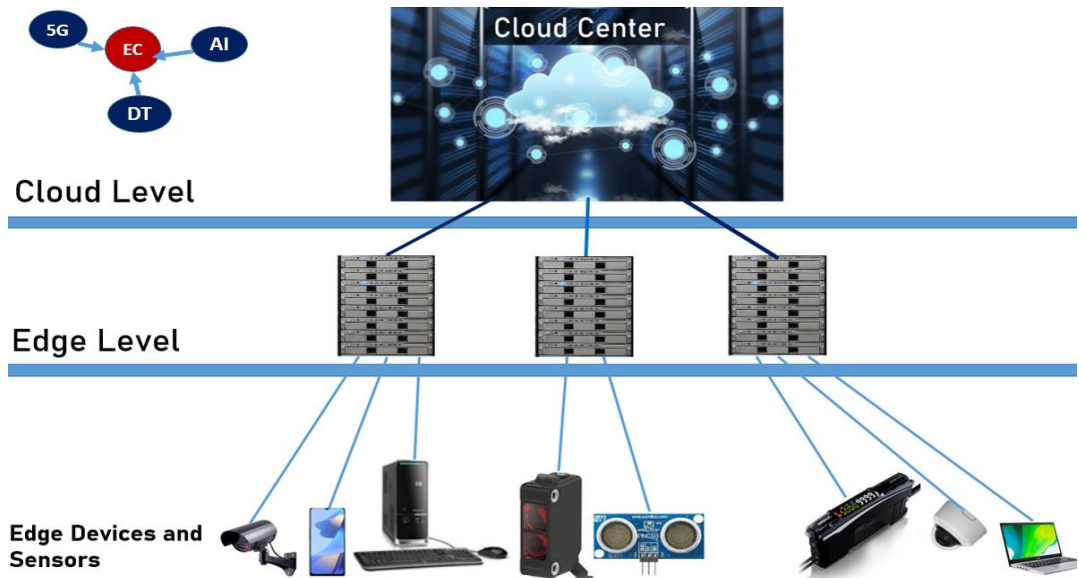
Figure 1. Edge-cloud computing architecture.

Edge computing stands at the intersection of 5G, artificial intelligence (AI) and digital transformation (DT).

Edge-cloud architecture eliminates the necessity of allocating individual bandwidth for each device on the main cloud. This significantly optimizes the overall computing power available on the entire network by executing mission critical tasks at the edge level. Introducing the additional layer solves a series of issues by adopting the so-called *data minimization* principle, according to which the amount of data being sent to the main cloud should be minimized while keeping the whole chain of the data under control.

Nowadays, edge-cloud architecture stands for a good platform for maintaining real-time information systems which impose deadlines for producing control responses. For such systems, provision of proper transactions depends not only on their logical correctness, but also on operability in the required time. Few examples of timely computation environments are: urgent medical diagnosis; nuclear plants management; traffic control, real-time speech and image processing; video surveillance, etc. Traditional cloud technologies do not address timely requirements for data processing.

In edge computing, data are collected, processed and analysed hierarchically at different layers on the network according to a predefined role distribution scheme, still near the source where they are generated. Thus, besides data collection, the edge level may implement timely intelligent tasks, such as:
- data aggregation, classification and validation,
- peer-to-peer communication among regrouped devices (if needed),
- visualization,
- data filtering,
- distribution, dissemination and replication,
- data mapping and reducing,
- caching, chaining, backup,
- data and cryptographic key splitting, etc.

As for any information system, a major issue in edge computing is data security. With many benefits of edge computing, including higher efficiency and lower latency, data generated and stored locally on IoT devices are relatively easy to access due to their lack of cyber protection.

The remainder of this paper is organized as follows. Section 2 presents a review of related works in the field. Section 3 presents the proposed mechanism for edge-cloud hierarchical data construction. The suggested confidential computing is given in Section 4 including hierarchical data linkage, the blockchain, cryptographic key management and related topics. A brief Summary of the paper is given in Section 5.

## 2. RELATED WORKS

Timely requirements affect data intelligence related to methods and tools for integrating and classifying real-time information collected from IoT smart devices, and, as authors in [1,2] state, given the unpredictable nature of the devices, maintaining and securing fast and high-quality algorithmic results in the face of uncertainty is still a challenging problem. The authors propose to tackle this challenge by applying coding theoretic techniques to provide resiliency against noise.

Edge/Fog computing [3] emerges as a novel computing paradigm that harnesses resources in the proximity of the IoT devices so that, alongside with the cloud servers, provides services in a timely manner. In the work, the authors proposed a weighted cost model to minimize the execution time and energy consumption of IoT applications in a computing environment with ever-increasing growth of IoT devices, multiple fog/edge servers and cloud servers. The paper introduces a new application placement technique based on the Memetic Algorithm to cover concurrency of IoT applications. In order to address IoT applications' heterogeneity, a lightweight pre-scheduling algorithm to maximize the number of parallel tasks for concurrent execution has been suggested.

The work [4] introduces a survey on multi-access edge computing architecture and proposes an elaboration of the cloud computing platform via the deployment of storage and computational resources at the edge thus reducing latency of edge devices and utilizing the core network more efficiently.

Authors in [5] rightly state that cloud and edge/fog computing are non-interchangeable technologies, and they cannot replace one another, but, when combined, they can contribute to promoting hierarchical data infrastructure and layered data processing.

In [6], authors interpret fog/edge computing-based IoT to be the future infrastructure on IoT development and an important computing paradigm in realizing the intelligent cyber-physical world.

Authors of the paper [7] highlighted typical security issues in edge architecture. Besides, they explored security algorithms acceptable for governing IoT data privacy taking into account the resource-constraint nature of the devices. Particularly, elliptic curve based cryptographic solutions were selected as most appropriate.

Understanding and implementing edge computing implies consideration of traditional computing paradigms [8], exploration of which can orientate the developers in selecting and combining relevant approaches. Besides the well-known methodologies in computing, trusty systems' development requires embedding confidential computing [9,10] which preserves the data confidentiality not only in the rest and in transfer, but also in use. Shielding the sensitive data, also the whole set of applied programming techniques from the rest of the entire system, makes the data being processed to become accessible solely to authorized programming codes. As authors in [11] state, driven by the need to protect computations delegated to co-tenanted machines operated by cloud computing services, instruction set architectures may introduce and maintain isolates offering strong integrity and confidentiality guarantees to code and data contained within.

## 3. EDGE-CLOUD HIERARCHICAL DATA MANAGEMENT

Edge computing paradigm can benefit from hierarchical data construction and management strategy by applying unsupervised learning for classifying the data being stored and processed on different layers on the entire network.

In the simplest implementation, we propose splitting data collected at edge level, as follows:
  a) leaving and processing timely data at the edge level, and
  b) transferring the historical data, also skills (successful or unsuccessful stories) gained through intelligent actions to the main cloud. Here, both successful and unsuccessful stories will be assimilated by the so-called "Knowledge Base" embedded system and will contribute to the overall system wisdom.

The following should be ensured prior splitting the data into disjoint segments:
  a) data distribution within specific groups of devices is the same, or
  b) data distribution changes over time.

In cases when data distribution is undefined, or probability density function is unknown (yet), investigation of data distribution scenarios prior to data splitting should be done in order to best regroup the devices with similar behaviour, such as: identical computations; area similar surveillance, etc. Appropriate machine learning tool, such as, determination of the best fitting data distribution (using Python) may be implemented in order to recognize data theoretical distributions. Given row data samples, X, the following code fragment may facilitate the data exploration process based on a Residual sum of squares (RSS) metric.

```
import numpy as np
from distfit import distfit
dist = distfit()              # initialize distfit
dist.fit_transform(X)         # determine the data best fitting probability distribution
print(dist.summary)
dist.plot()
```

Outliers detection is another important point in the recognition of data distribution. Variance and standard deviation detection will reveal new data points that deviate significantly. Those data points will then be marked and processed accordingly.

Data distributions revealed, appropriate intelligence can be successfully applied for data analysis and decision making.

This overall process resides at the edge of the edge-cloud network.

## 4. CONFIDENTIAL COMPUTING IN EDGE-CLOUD HIERARCHICAL NETWORK

Data sharing is the process of making the same data accessible to a number of users or applications based on the one-to-many principle of association.

Note that data within a device are under the sole control and are of a required physical and cyber security. In this sense, the edge level must ensure the confidentiality and integrity of data being shared or exchanged in order to exclude exposing data to unauthorized access.

In edge computing, some portions of data are shared with the main cloud for a deeper and probably latent analysis. The edge level may even split and share cryptographic keys, tokens and other cyber identifiers in order to keep trace of the shared data between computational and storage levels. Particularly, while splitting is done within a specific edge data point and under the local control, sharing data parts results in transferring them, and this is when solely GDPR compliant solutions have to be embedded in order to ensure zero-trust based identification and strong authentication of the shares. The amount of available and intelligible data consequently decreases and resides in the scope of confidential computing.

Privacy at the edge is of a great importance and implies preserving data confidentiality, integrity and availability. For this purpose, we suggest delegating most of the cryptographic operations to edge servers while embedding pre-image resistant and collision-free cryptographic hash functions.

A hash function $H: S \rightarrow S'$ is a mapping from a set of arbitrary cardinality ($S$) to a single value from a set of a fixed (fewer) cardinality ($S'$). Cryptographic one-way hash function $H$, with the domain $S$; the range $S'$; $s \in S$, and $y \in S'$, is of a polynomial time computation for all $s \in S$, while finding $s: H(s) = y$, is an NP-complete problem. This feature provides pre-image resistance of the function. If $Prob\_of\_finding\ (M, M'): H(M) = H(M') \rightarrow 0$, then the mapping $H$ is also collision-free (Russell 1992).

Besides, the following mechanisms are suggested to be embedded at the edge:
- Hardware based tamper-proof storage (HSMs) for
    - cryptographic key generation
    - data encryption
    - cryptographic key encryption
    - cryptographic key splitting
    - verifiable secret sharing
    - cryptographic key shares' combination and algebraic verification,
    - secure computations
- Honey Encryption combined with Advanced Encryption Standard algorithm for data encryption at rest and at transfer.

Identity management ensures automatic identification and authentication of edge devices and relevant edge computing resources. For this purpose, the following mechanisms should be embedded at the edge level.
- IoT Device Identity Lifecycle Management
- Authentication, in order to ensure validity of entities within the system. This is the first step to perform any computing operation before moving to the next steps
- Authorization, that always comes after the authentication and allows solely verified entities to get access to resources
- Distributed identity management architecture, equipped with 5G network, in order to assign unique identities by the software defined identity management systems. Here, local nodes will share parameters with the local Software Defined Network servers to acquire identity. Local SDN devices synchronize their identity databases with the main cloud SDN device. All Global SDN devices will also synchronize their databases. Thus, the whole hierarchy in the edge architecture will be strongly identified.

Whenever appropriate, timestamp may be embedded.
A timestamp is a mapping $T:(A \times L \times G) \rightarrow S$, where
- A is an alphabet,
- L is the set of literals,
- G is the global TSA (Time Stamp Authority, the global secure time stamp server) data. The TSA will register the current transaction date and time which, when necessary, can be verified with the main cloud to resist to replay attacks.
- S presents the resultant timestamp string.

In cases, when embedding PKI (Public Key Infrastructure) is expensive, self-certified cryptography may be utilized to implement the registration and at least two-factor authentication of network entities. Meanwhile, authentication is verified each time before edge and main cloud levels interact. Internally generated certificates can be bound to identities and construct the identity and certificate management mechanism based on blockchain.

It is well known that blockchain is a digitally distributed, decentralized, public ledger that exists across a network. No single entity controls the blockchain network; anyone can join at any time. The above premises provided, identification of IoT devices within a group and on the main cloud (when necessary) is achieved as described below.

Blockchain identity management systems are commonly used to address identity issues such as data insecurity and fraudulent identities. The distributed ledger, blockchain, will store:
- ✓ complete paths (individual or joint, depending on the implementation) for each device having data shared between the edge and main cloud,
- ✓ labelling of the classified data at their cascading aggregation, validation, classification and then filtering phases, in order to trace the data hierarchical chain from the edge to the main cloud.

Due to resource-constraint nature of IoT devices, we propose constructing of the edge-to-main cloud blockchain based on secure computations on elliptic curves. Unlike the simplified curves used in traditional data encryption and decryption, the selection of unique points here is done on an original curve in order to meet higher security requirements.

An elliptic curve E over the real numbers R is defined by a Weierstrass equation,

$$E: y^2 + a_1xy + a_3xy + a_3y = x^3 + a_2x^2 + a_4x + a^5$$

with coefficients $a_1, a_2, a_3, a_4, a_3 \in N$, and $\Delta \neq 0$ discriminant.

The set of points on the curve is:

$$E(L) = \{(x, y) \in R \times R : y^2 + a_1xy + a_3xy + a_3y = x^3 + a_2x^2 + a_4x + a^5 = 0)\} \cup \{0\}$$

with the point of infinity (the 0 point).

Initially, prior regrouping the devices, the ledger is an empty list. The list gets incrementally updated with every new device joining the specific group and presents a dynamic list of unique identifiers obtained in the following steps:

a) For every device in a group, a unique point on the elliptic curve is selected and recorded at the edge secure server zone. AES 256 is applied to encrypt the points according to data anonymization principle: *encrypted data is neither usable nor decrypted* as they are not used in any of further transactions. Utilizing elliptic curves is motivated by an NP-completeness of guessing the point coordinates even if the curve is made public.

b) The encrypted point is hashed with the timestamp $H_1$ *(Point, Timestamp)* using SHA256.

c) $H_1$ *(Point, Timestamp)* is hashed with the previous content in the public ledger (this step is skipped with the first record).
The resultant $H_2$ *(Point, Timestamp, History)* is the unique ID for a device joining the group.

d) $H_2$ is recorded in the edge ledger.

e) The ledger is transmitted to the main cloud server to a distinct location.

Distinct paths for each blockchain provided, anyway, access control mechanisms that meet the edge computing security, privacy and data diversity requirements should be upgraded from traditional access control schemes. For this purpose, involving Bloom filter integrated with identity management and lightweight secret key agreement protocols based on self-authenticated public key may serve as a good basis for innovating edge/cloud access control.

Cryptographic key management at the edge administers the full lifecycle of hierarchic cryptographic keys.

It is strongly recommended to limit
- ✓ the amount of information protected by a given key,
- ✓ the amount of exposure if a single key is compromised,

Note, that edge-level encryption secures the data collected from edge devices such that no level in computing receives the raw version of the payload directly. For effective and secure key management, HSM and related secure zone at the edge level should be embedded. This will ensure supporting sensitive operations, like:
- ✓ key generation, key secure storage and key encryption,
- ✓ key distribution, backup,
- ✓ verifiable secret splitting and sharing, etc.

In order to support two-level cryptographic key sharing, the edge-level HSM internally performs:
- ✓ cryptographic key generation,
- ✓ cryptographic key encryption applied AES-256,
- ✓ cryptographic key verifiable secret splitting. One portion of the encrypted key remains at the edge secure zone, while the second portion is transferred to the main cloud
- ✓ each time a transaction is activated on the main cloud site, firstly a verification of the validity of the key shares is performed.

Most of existing secret sharing schemes of proven security (like in Shamir's threshold scheme) are based on the assumption that all participating users are legitimate. This approach is prone to sophisticated attacks: the attacker can impersonate a legitimate party without being detected. Schemes of proven security can be found in [13,14] where the impersonation attack is advised to be blocked based on modification of Shamir's threshold scheme, or based on plane parametric curves with one-parameter representation for a master key, respectively. Other solutions are arising from Latin squares [15].

We propose constructing a verifiable secret sharing scheme based on abstract structures from non-associative algebras. The theory of quasigroups is a permanently evolving scientific direction. Quasigroups are based on the Latin square property and stand for a generalization of groups without the associative law or identity element [16].

The attractiveness of quasigroups in construction of verifiable secret splitting and sharing schemes is in their easily programmable nature due to utilization of solely logical operations. A quasigroup with its parastrophs $(Q, \cdot, \backslash, /)$ is a set closed under three different binary operations, referred to as multiplication $(\cdot)$, left division $(\backslash)$ and right division $(/)$ satisfying the conditions:

1. $x \cdot (x \backslash y) = y$
2. $(y / x) \cdot x = y$
3. $x \backslash (x \cdot y) = y$
4. $(y \cdot x) / x = y$
5. $x / (y \backslash x) = y$
6. $(x / y) \backslash x = y$

The verifiable secret splitting is performed as follows:
1. the order *n* of the quasigroup (the number of its elements) dictates the number of shares
2. edge computing paradigm suggests having (2-out-of $2^n$) shares satisfying above properties for the secret quasigroup
3. the two shares are encrypted and distributed between the network layers
4. when combined, the two shares are decrypted within the edge level HSM, where shares are authenticated.

The proposed scheme eliminates the risk of impersonation attacks against both the edge and the main cloud levels: the shares are verified at the edge level HSM by a polynomial time computation (meanwhile for the non-legitimate party this computation will lead to a numerical explosion with a large order of the secret quasigroup), and the main cloud site is secured by confidential computing.

Another significant factor motivating the usage of quasigroups is that generalized identities of higher order logics can be effectively constructed on quasigroups without having any significant impact on algorithmic complexity.

Confidential computing assumes hardware-level isolation of data and data processing in a tamper proof and trusted execution environment (TEE) aimed at preserving:
  a) data confidentiality: unauthorized entities cannot view data while it is in use,
  b) data integrity: unauthorized entities cannot add, remove, or alter data while it is in use,
  c) code integrity: unauthorized entities cannot add, remove, or alter code executing in the TEE

While encryption helps in providing cybersecurity for data in rest and in transfer, confidential computing solves the remaining data vulnerability and enables to trust the code instead of the computing system physical security. Confidential computing isolates the data from attackers, insiders and even from the infrastructure operators via keeping the data and data processing in enclaves of the operating system and machine memory.

Unlike cryptographic keys generated and stored in software which aren't hard to find, sandboxed enclaves developed for confidential computing safeguard data in use.

## 5. CONCLUSIONS

The paper presents a series of approaches for implementing confidential computing within edge-cloud hierarchy. The suggested method for classifying IoT devices into separate groups are introduced and based on exploration of data distribution. Data splitting, chaining, sharing and logical linkage are achieved by embedding blockchain enforced with verifiable secret sharing. Tamper-proof enclaves ensure data privacy in their use.

**Authors**

Yeghisabet Alaverdyan. Head of Systems Integration Department at EKENG CJSC since 2018. Graduate of the National Polytechnique University of Armenia. PhD, Associate Professor. Author of more than 25 scientific papers published in local and international scientific journals. Research is in Mathematical Cryptography, AI and Cognitive systems modelling.

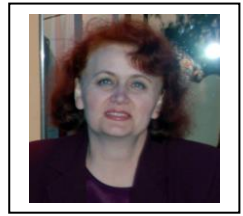

Suren Poghosyan. Lead of Scientific group at the IIAP NAS RA since 1988. Graduate of the Yerevan State University. Author of more than 25 scientific papers published in local and international scientific journals. PhD. Research is in Self-organized systems modelling.

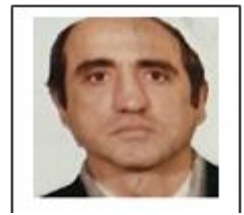

Vahag Poghosyan. Research fellow at the IIAP AS RA and Senior Software developer at Synopsis Armenia. Graduate of the Yerevan State University. PhD. Author of more than 30 scientific papers published in local and international scientific journals. Research is in Self-organized systems modelling.

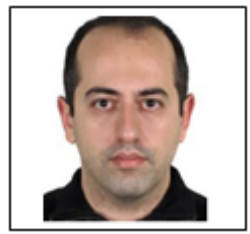